\documentclass[%
 reprint,
 amsmath,amssymb,
 aps,
]{revtex4-2}

\usepackage{graphicx}
\usepackage{dcolumn}
\usepackage{bm}

\usepackage{filecontents}

\begin{document}
\preprint{APS/123-QED}

\title{Ultrafast generation of hidden phases \\ via energy-tuned electronic photoexcitation in magnetite}
\author{B. Truc}
\affiliation{Institute of Physics, LUMES, École Polytechnique Fédérale de Lausanne (EPFL), Lausanne CH‐1015, Switzerland}
\author{P. Usai}
\affiliation{Institute of Physics, LUMES, École Polytechnique Fédérale de Lausanne (EPFL), Lausanne CH‐1015, Switzerland}
\author{F. Pennacchio}
\affiliation{Institute of Physics, LUMES, École Polytechnique Fédérale de Lausanne (EPFL), Lausanne CH‐1015, Switzerland}
\author{G. Berruto}
\affiliation{Institute of Physics, LUMES, École Polytechnique Fédérale de Lausanne (EPFL), Lausanne CH‐1015, Switzerland}
\author{R.Claude}
\affiliation{Institute of Physics, LUMES, École Polytechnique Fédérale de Lausanne (EPFL), Lausanne CH‐1015, Switzerland}
\author{I. Madan}
\affiliation{Institute of Physics, LUMES, École Polytechnique Fédérale de Lausanne (EPFL), Lausanne CH‐1015, Switzerland}
\author{V. Sala}
\affiliation{Dipartimento di Fisica, Politecnico di Milano, Piazza Leonardo da Vinci 32, Milano, Italy}
\author{T. LaGrange}
\affiliation{Institute of Physics, LUMES, École Polytechnique Fédérale de Lausanne (EPFL), Lausanne CH‐1015, Switzerland}
\author{G. M. Vanacore}
\altaffiliation[Currently at ]{Department of Materials Science, LUMiNaD, University of Milano-Bicocca, Via Cozzi 55, 20125 Milan, Italy}
\affiliation{Institute of Physics, LUMES, École Polytechnique Fédérale de Lausanne (EPFL), Lausanne CH‐1015, Switzerland}
\author{S. Benhabib}
\email{siham.benhabib@epfl.ch}
\affiliation{Institute of Physics, LUMES, École Polytechnique Fédérale de Lausanne (EPFL), Lausanne CH‐1015, Switzerland}
\author{F. Carbone}
\email{fabrizio.carbone@epfl.ch}
\affiliation{Institute of Physics, LUMES, École Polytechnique Fédérale de Lausanne (EPFL), Lausanne CH‐1015, Switzerland}

\date{\today}

\begin{abstract}
Metal-insulator transitions (MIT) occurring in non-adiabatic conditions can evolve through high-energy intermediate states that are difficult to observe and control via static methods. By monitoring the out-of-equilibrium structural dynamics of a magnetite (Fe$_3$O$_4$) crystal via ultrafast electron diffraction, we show that MITs can evolve through different pathways by properly selecting the electronic excitation with light. Near-infrared (800\,nm) photons inducing d-d electronic transitions is found to favor the destruction of the long-range zigzag network of the trimerons and to generate a phase separation between cubic-metallic and monoclinic-insulating regions. Instead, visible light (400\,nm) further promotes the long-range order of the trimerons by stabilizing the charge density wave fluctuations through the excitation of the oxygen 2p to iron 3d charge transfer and, thus, fosters a reinforcement of the monoclinic insulating phase. Our experiments demonstrate that tailored light pulses can drive strongly correlated materials into different hidden phases, influencing the lifetime and emergent properties of the intermediate states.
\end{abstract}

\maketitle

\section{\label{sec:level1}Introduction}
The physical properties of strongly correlated materials are mainly defined by the complex interplay among the electronic, orbital, spin, and atomic degrees of freedom. At equilibrium, phase transitions follow an ergodic pathway within the material free-energy landscape, and the transition is characterized by a succession of thermodynamic equilibrium states between two global minima. Instead, using ultrashort laser pulses drive the transition out-of-equilibrium and induce a distinct pathway by transiently changing the coupling between the relevant degrees of freedom. Light-driven phase transitions reveal the presence of new intermediate (hidden) states of matter \cite{Fausti2011,Kiryukhin1997,Stojchevska2014,Miyano1997,Ichikawa2011,Gao2022,Zakery2021,Koshihara1990,DelaTorre2021}. Such hidden phases are not only of interest from a fundamental point of view but also bear potential for ultrafast technological devices \cite{DelaTorre2021,Giustino2020}.\\
Magnetite (Fe$_3$O$_4$) is a prototypical strongly-correlated system. It exhibits a complex interplay between the crystal structure \cite{Iizumi1982}, charge \cite{Nazarenko2006,Goff2005,Lorenzo2008,Wright2001} and orbital orders \cite{Leonov2004,Jeng2004}, which leads to the emergence of an atypical thermodynamic MIT in the vicinity of 125\,K, known as the Verwey transition (VT) \cite{VERWEY1939}. It is found that the structural changes play a key role in VT \cite{Ihle1980,Yamada1975}. Above Verwey temperature (T$_{V}$), magnetite has a cubic inverse spinel $Fd\bar{3}m$ structure formally written Fe$^{+3}$[Fe$^{+2}$Fe$^{+3}$]O$_{4}$, the first Fe$^{+3}$ (A-type) occupied the tetrahedral sites, whereas the [Fe$^{+2}$Fe$^{+3}$] (B-type) occupied the octahedral sites.  Below the T$_{V}$, the symmetry changes from the cubic $Fd\bar{3}m$ to the monoclinic $Cc$ phase \cite{Iizumi1982}.\\
In the low-temperature (LT) phase, a new kind of bond dimerized state, the so-called trimeron, has been discovered and shown to form a long-range order \cite{Senn2012,Senn2012a}. The trimeron unit results from multiple cooperative effects, including charge, $t_{2g}$ orbital orderings, and strong electron-phonon coupling \cite{Piekarz2021}. Therefore, trimerons are deemed to be the key actor of VT, which has been recently described microscopically as an order-disorder transition from a trimeron liquid with incommensurate fluctuations to a commensurate crystal below T$_{V}$ \cite{Borroni2017a,Borroni2020,Borroni2017}. Optical experiments have shown that light offers the intriguing possibility to manipulate such charge fluctuations resulting in tuning the electron-phonon coupling \cite{Borroni2017} and suggesting that the light-induced transition can be very orbital-selective.\
Here, we directly visualize the out-of-equilibrium structural dynamics of a magnetite single crystal employing ultrafast electron diffraction (UED).\\
UED allows us to track the lattice evolution of magnetite across the photoinduced MIT. We show that, depending on the photon energy of the femtosecond optical excitation used in the experiment (1.55\,eV vs. 3.10\,eV), we trigger different electronic excitations, consequently leading to distinct nonequilibrium metastable structural states. 
\section{\label{sec:level2}Experimental Methods}
The UED experiments were performed in reflection geometry \cite{Mancini2012,Pennacchio2017} with a grazing angle of 0.5$^{\circ}$ to 5$^{\circ}$. The light source is a Ti:sapphire laser amplifier with a central wavelength of 800\,nm with a pulse duration of  45\,fs at a repetition rate of 20\,kHz. 
 
High-quality magnetite with T$_{V}\approx$ 117\,K exposing the flat and optically polished (110) surface was fixed on a cold finger attached to a five-axis manipulator with silver conductive paste and placed inside an ultrahigh vacuum chamber ($\leq$ 10$^{-9}$\,mbar). The temperature range is controlled using an open cycle cryostat with helium liquid flow (see the SI).   

\section{\label{sec:level3}Results}
 We have first monitored the quasi-adiabatic transition induced by varying the temperature of magnetite (without photoexcitation).  We investigate the structural changes by decreasing the temperature from 150\,K down to 40\,K and simultaneously following the quasi-static change of the diffraction pattern along the [110] direction where anomalies attributed to the trimerons have been recently observed \cite{Lin2015}.\
\begin{figure}[t!] 
\centering 
\includegraphics[width=1\columnwidth]{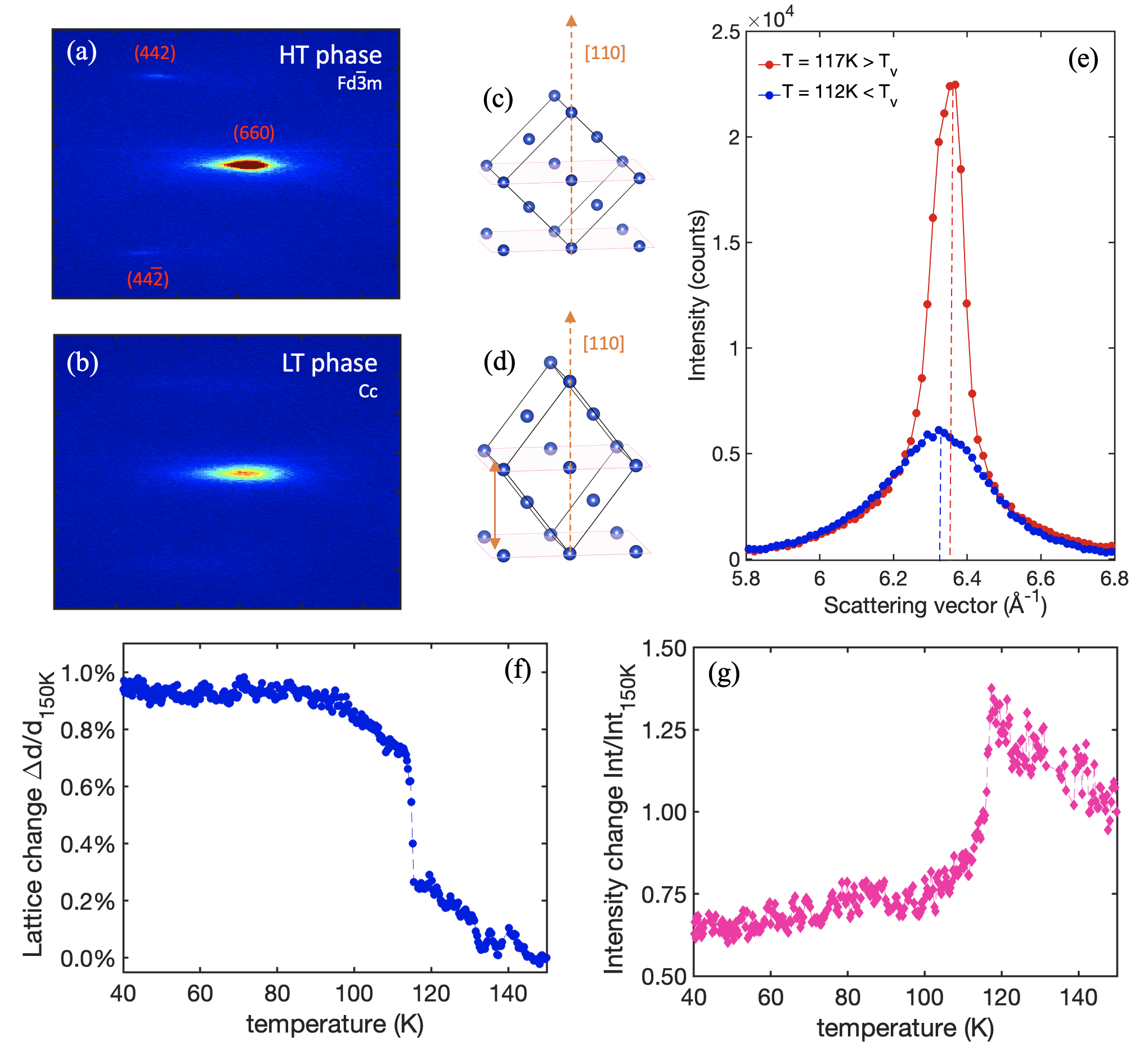}
\caption{Static diffraction pattern of magnetite in the cubic phase $Fd\bar{3}m$ above T$_{V}$, T=150\,K (a) and in the monoclinic phase $Cc$ below T$_{V}$ at T=40\,K  (b). (c) Cubic face-centered structure oriented along the [110] direction reflecting the cubic unit cell of magnetite above T$_{V}$. (d) The elongated unit cell of magnetite below T$_{V}$. (e) Peak profile of the (660) Bragg peak above and below T$_{V}$, its parameters are extracted from a Voigt profile fit. (f) Temperature dependence of the interplanar atomic change (expansion/compression) along [110] extracted from the (660) Bragg peak position (see SI). (h) Evolution of the integrated area under the (660) Bragg peak across the transition. }
\label{fig:1} 
\end{figure}

In Fig.\ref{fig:1}(a) and (b), we show the static diffraction patterns of magnetite measured above and below T$_{V}$, respectively. Specifically, we monitor the changes of the (660) Bragg peak of the cubic phase and observe a significant modification of the peak position from which we extract the atomic interplanar expansion or compression (see SI) shown in Fig.\ref{fig:1}(f). This is accompanied by a drop in the diffraction intensity as illustrated in Fig.\ref{fig:1}(g). Such intensity drop results from the combination of two concomitant factors: i) the lowering of the structure factor when transitioning from a high-symmetry cubic phase to a lower symmetry monoclinic structure, and ii) the incoherent electron scattering process through multiple micro-sized structural domains (twins) that emerge in LT phase \cite{Kasama2010}. Across VT, at 117\,K, the cubic lattice transforms into the monoclinic phase, which is evidenced by the expansion of the lattice along the [110] direction, as sketched in Fig.\ref{fig:1}(d). This specific expansion significantly changes the shear strain $\varepsilon_{xy}$ (see  SI). In addition, it causes a significant softening in the shear elastic constant $c_{44}$, as reported in ultrasound measurements \cite{Schwenk2000}. Although our technique is moderately surface sensitive (5-10\,nm), the agreement between the monitored position shift of the Bragg peak from our data and the reported softening in the elastic constant $c_{44}$ demonstrates that our observations are representative of bulk dynamics.\\
Ginzburg-Landau’s (GL) theory of phase transition such as VT in magnetite correlates transformation shear strains to an order parameter. Hence, we ascribe the measured shear strain ($\varepsilon_{xy}$) being strongly coupled to the order parameter (OP) $\Delta$, and retrieve its symmetry based on the framework of GL \cite{Landau1980} and a fundamental group theory analysis (see SI). We found that $\Delta$ has T$_{2g}$ symmetry with one nonzero component, i.e., $\Delta$ = ($\Delta_{xy}$, 0, 0).\\
Using detailed group theory calculations, several authors have identified the set of phonons, including $\Delta_5$, $X_3$, and $\Gamma^{+5}$  (T$_{2g}$), as the structural OPs \cite{Piekarz2007,Gasparov2000,Piekarz2006}. Furthermore, \textit{ab initio} calculations have demonstrated the strong coupling between these structural OPs and the T$_{2g}$ orbital ordering within a trimeron \cite{Piekarz2007}. Therefore, we conclude that the trimerons arrangement along the [110] direction is a conceivable OP candidate with T$_{2g}$ representation. The OP with T$_{2g}$ representation was suggested recently by electron diffraction measurements, where the authors consider an anomalous electronic nematic phase above Tv with a T$_{2g}$ representation which involves a different set of rotational symmetry breaking than the usual ones \cite{Wang2022}.

\begin{figure*}[t!] 
\centering 
\includegraphics[width=1\textwidth]{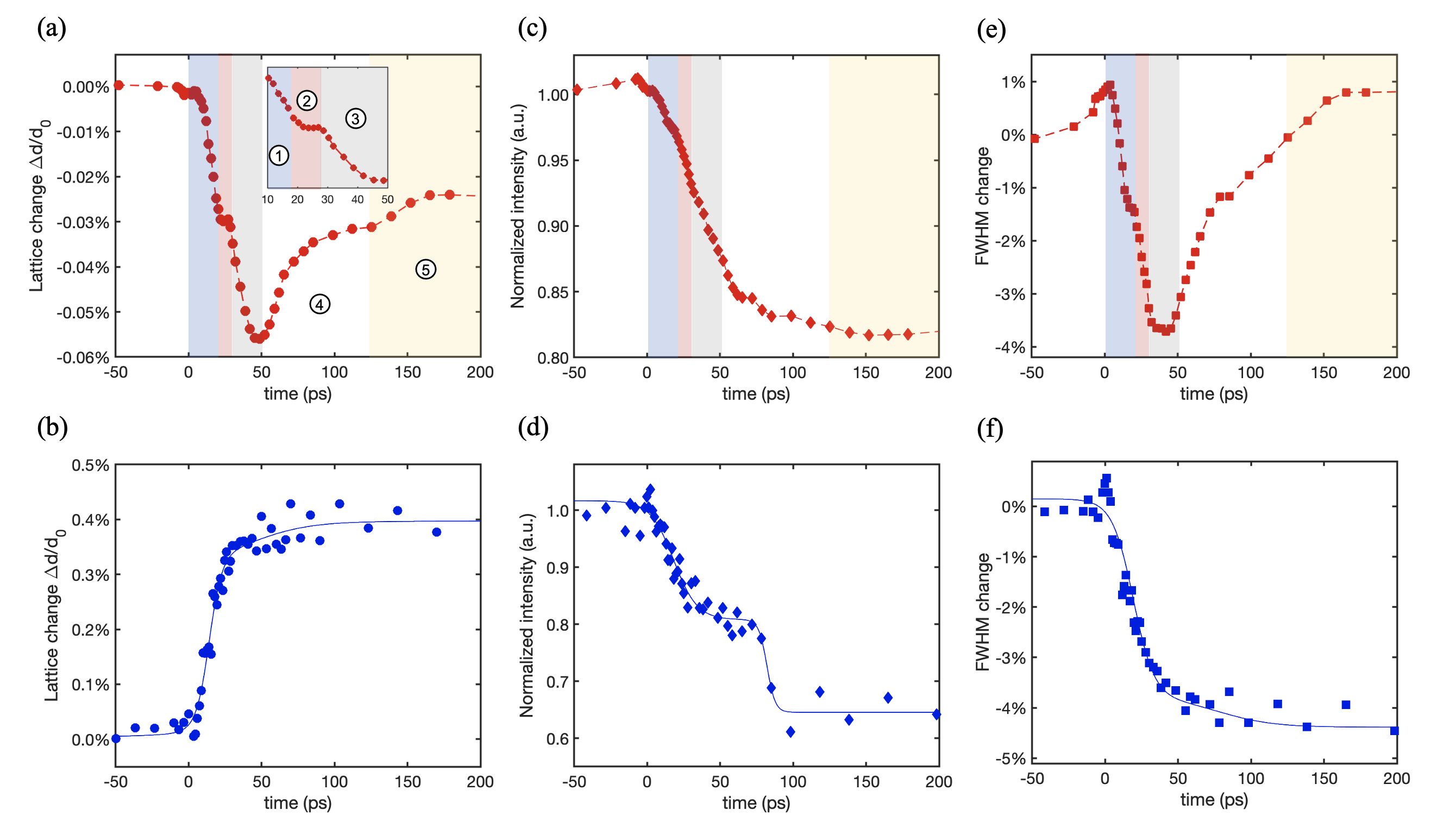}
\caption{Evolution of the lattice change compression/expansion along the [110] direction (a) under 800\,nm and (b) 400\,nm photoexcitation. In (a), the shaded areas show multiple compression stages. (c) and (d) evolution of the normalized intensity under 800\,nm and 400\,nm photoexcitation, respectively. (e) and (f) time dependence of the FWHM (full width at half maximum) under 800\,nm and 400\,nm photoexcitation, respectively. Solid lines are guides to the eye.}
\label{fig:2} 
\end{figure*}

We obtain insights into the non-adiabatic MIT in magnetite by investigating the out-of-equilibrium response of the lattice, initially kept at 80\,K (below T$_V$), after photoexcitation. In Fig.\ref{fig:2}, we present the evolution of the (660) Bragg peak following ultrashort laser pulses at two different wavelengths, 800\,nm and 400\,nm at 2.9\,mJ/cm$^2$ and 1.2\,mJ/cm$^2$ incident fluence, respectively. The fluence used for 800\,nm corresponds to the \textit{intermediate} fluence regime \cite{Randi2016}.\

Fig.\ref{fig:2}(a) shows the structural dynamics following 800\,nm photoexcitation. The (660) atomic planes undergo a maximum compression of around -0.06\,\%. Based on our static data shown above (Fig.\ref{fig:1}), the compression of the monoclinic lattice along the [110] direction denotes the transformation toward a cubic structure, consistent with ref.\cite{Randi2016}. Their recent out-of-equilibrium optical measurements have shown a photoinduced phase separation between insulating regions and metallic islands at LT through 800\,nm laser pulses in a similar fluence regime. This observation is supported by pump-probe x-ray diffraction measurements \cite{DeJong2013}, where the authors attribute the insulating state to the monoclinic regions and the metallic state to the cubic islands. Expanding the investigation range of previous measurements that were limited to the first 10\,ps \cite{Wang2022,Randi2016,DeJong2013}, our data unveil the complex establishment of the hidden phase which lasts approximately 50\,ps and interestingly follows three compression stages. First, during the first 22\,ps (N°1), an abrupt compression of -0.03\,\% occurs. In the second stage (N°2) between 22\,ps and 27\,ps, the lattice undergoes a minor contraction. Finally, the third step (N°3) emerges and adds an extra -0.03\,\% to the lattice compression. This multi-step process is characteristic of the presence of distinct dynamic processes, such as electron-phonon coupling and phonon-phonon interaction \cite{Maldonado2017,Wang2022}. Note that the electron-electron interaction is expected to occur on a faster timescale $<\,$300\,fs \cite{DeJong2013}, beyond the temporal resolution in these experiments. The relaxation process also evolves with multiple timescales. Qualitatively, the first recovery stage (N°4) occurs from 50\,ps to 126\,ps and reaches an intermediate compressed state close to -0.03\,\%, which interestingly corresponds to the value of the process N°3. Then, a second long process occurs towards the total recovery (N°5) to the equilibrium phase, which is still not reached after 1.3\,ns (see SI). In a second data set with a slightly higher fluence (see SI), we confirm the multiple compression stage process and show that each stage's duration and amplitude depend on the fluence used.

In Fig.\ref{fig:2}(c), the response of the Bragg peak intensity after the 800\,nm photoexcitation shows a drop. Taking only into account the recovery towards the higher symmetry cubic high temperature phase after the 800\,nm photoexcitation, we expect an increase in the intensity, which we do not observe. This observation suggests that the dominant process is the structural disorder caused by the motion of the atoms induced by the rise in the lattice temperature by ultrashort laser pulses \cite{VanTendeloo2011}, known as the induced Debye-Waller effect (see SI). Similarly to the peak position shift response, the Bragg peak intensity does not fully recover to its initial state after 1.3\,ns (see SI). The long dynamics revealed by our data indicate the metastability of the induced 800\,nm phase illustrating the complexity of the thermalization process in magnetite, involving multiple interplays of electron-electron, electron-phonon, and phonon-phonon scattering. Such lifetime is a signature of a hidden out-of-equilibrium phase which is supported by the unusual structural dynamics and the decrease in the intensity due to the phase separation between cubic and monoclinic islands, which is not an equilibrium state but rather a local minimum within the energy landscape of the excited configuration. 

We extend our investigation for hidden phases in magnetite by changing the energy of the optical excitation to 3.10\,eV (400\,nm). In Fig.\ref{fig:2}(b), different from the 800\,nm case, we observe that the  400\,nm laser pulses induce a 0.4\,\% expansion of the lattice along the [110] direction (instead of a contraction), indicating a reinforcement of the monoclinic distortion. At 90\,K, before excitation, the crystalline structure has a monoclinic angle $\beta_{M}$= 90.236$^{\circ}$ \cite{Senn2012}. The 400\,nm induced expansion is mainly related to the variation of the tilting angle $\beta_{M}$, where we expect a value $\beta_{M400\,nm}$ $>$ 90.236$^{\circ}$. A quantitative value for the tilting angle can only be retrieved when monitoring the behavior of multiple Bragg peaks along different zone axes. Nevertheless, our data clearly show that the 400\,nm optical excitation induces a lattice change that is opposite to the quasi-adiabatic lattice response from our equilibrium data presented above (Fig.\ref{fig:1}(f)), where the stabilization of the structure from 90\,K down to 40\,K shows no modification of the lattice parameters expected thermodynamically. Since the generated structure is not accessible thermally but only induced optically, we associate it with the emergence of a new hidden phase characterized by a monoclinic lattice with a tilting angle larger than the equilibrium value of 90.236$^{\circ}$. The 400\,nm hidden phase is also completely different from the one established by the 800\,nm light. The first is firmly monoclinic, whereas the second is a mixture of monoclinic and cubic separated regions. The 400\,nm structural hidden phase takes around 50\,ps to emerge with only one direct expansion process, as presented in Fig.\ref{fig:2}(b), which we relate mainly to electron-phonon interaction. This new state lasts up to 300\,ps without any recovery to the initial state giving it a metastable character. We observe a significant drop in the intensity response (Fig.\ref{fig:2}(d)). For the 800\,nm case, thermal effects and multiple scatterings from the mixed phase are the origins of the intensity drop. Although the decrease in intensity is consistent with the reduction of the structure factor for the (660) Bragg peak, it is surprising to observe a shrinking of the FWHM, indicating a higher homogeneity in the atomic planes, which we cannot associate with a thermal-like behavior. This suggests that the new hidden structural state possesses a larger structural long-range order related to a larger monoclinic angle.

\section{\label{sec:level4}Discussion}
The formation of distinct metastable hidden phases through two different photon energies demonstrates the critical role played by electronic excitations in establishing such nonequilibrium phases in magnetite. At LT, magnetite is thermodynamically stabilized in the insulating phase, resulting from a commensurate long-range order along the [001] direction \cite{Senn2012,DeJong2013} of the trimerons zigzag network \cite{Senn2012a,Reznicek2015} with a coherent length of (385±10)\,nm \cite{Lorenzo2008}. Each trimeron unit couples linearly three Fe$_\mathrm{B}$ sites in the form Fe$_\mathrm{B}^{3+}$\,-\,Fe$_\mathrm{B}^{2+}$\,-\,Fe$_\mathrm{B}^{3+}$, in which the minority spin t$_{2g}$ electron is delocalized from the central Fe$_\mathrm{B}^{2+}$ site into the nearest neighbors Fe$_\mathrm{B}^{3+}$, and each Fe$_\mathrm{B}^{3+}$ site is shared between three trimerons. In addition to the Jahn-Teller distortion caused by the orbital ordering \cite{Huang2017}, the electron localization within a trimeron unit produces a structural distortion in the Fe$_\mathrm{B}$ sites, where the distance between the central Fe$_\mathrm{B}^{2+}$ and its two Fe$_\mathrm{B}^{3+}$ nearest neighbors gets shorter, leading to a monoclinic distortion \cite{Senn2012}. This distortion has been confirmed by high-accuracy synchrotron x-ray structure refinements \cite{Senn2015}, where it is found that fourteen over the sixteen nonequivalent trimerons have shown a shorter Fe$_\mathrm{B}$ - Fe$_\mathrm{B}$\ distance. When we excite magnetite using photon pulses with an energy of 1.55\,eV (800\,nm), LDA+U calculations \cite{Leonov2004} and optical conductivity measurements \cite{Randi2016,Park1998} both agree in predicting the triggering of electronic d-d excitations. They correspond to an electron delocalization from an occupied t$_{2g}$ of Fe$_\mathrm{B}^{2+}$ to an unoccupied t$_{2g}$ orbital of Fe$_\mathrm{B}^{3+}$ following the configuration 3d$_{i}^6$3d$_{j}^5$ $\rightarrow$ 3d$_{i}^5$3d$_{j}^6$. The d-d excitations encompass another inter-site excitation corresponding to the transition from an occupied t$_{2g}$ of Fe$_\mathrm{B}^{2+}$ to an unoccupied e$_g$ of  Fe$_\mathrm{B}^{3+}$ \cite{Leonov2004}. However, this excitation is only possible at an energy higher than 2\,eV. Triggering the d-d excitation restores the mobility of the minority spin t$_{2g}$ electrons and causes the valency to change for both Fe$_\mathrm{B}^{2+}$ and Fe$_\mathrm{B}^{3+}$ and, hence, alternates their sites inside the trimeron (see SI). The direct consequence of this local electronic fluctuation is the destruction of the trimeron. According to out-of-equilibrium x-rays measurements, this destruction occurs in an ultrashort timescale $<\,300$\,fs \cite{DeJong2013}. The destruction of trimerons yields the suppression of the long-range zigzag order connected at Fe$_\mathrm{B}^{3+}$ sites. When the trimeron breaks, the Fe$_\mathrm{B}$ - Fe$_\mathrm{B}$\ distance returns to its initial value. Hence, the structure relaxes, giving rise to the emergence of cubic phase islands inside the remaining monoclinic regions forming a phase separation that we qualify as an 800\,nm metastable hidden phase. Following the electronic excitation, the relaxation process back to the equilibrium configuration continues via electron-phonon and phonon-phonon scattering mechanisms. Electrons primarily interact with high-energy optical phonons, which then anharmonically decay toward acoustic modes via a three-phonon scattering process \cite{Perfetti2007, Carbone2008, Mansart2013,Mansart2012a}. For magnetite, we can interpret the newly observed two steps in the relaxation process as the strong electron-phonon coupling involving the conduction electrons coupling with the X$_3$-driven mode (TO) for the first stage and the phonon-phonon coupling between X$_3$ (TO) and $\Delta_5$ (TA) modes for the second. This scenario is supported by recent UED data that have shown under similar photoexcitation (800\,nm) that the X$_3$ (TO) mode is preferably triggered via the electron-phonon coupling \cite{Wang2022} in the first few picoseconds. In addition, inelastic neutron scattering has shown that the $\Delta_5$ (TA) mode is the most susceptible across the cubic-monoclinic transformation \cite{Borroni2017a}.
 
In the equilibrium state, refined x-rays measurements show that in a high connectivity configuration with a maximum of three trimerons per Fe$_\mathrm{B}^{3+}$ site, only four of the eight Fe$_\mathrm{B}^{3+}$ sites participate, keeping the rest as inactive \cite{Senn2012}. In addition, our static data in Fig.\ref{fig:1} show that this thermodynamic maximum connectivity state is reached a few kelvins after T$_{V}$, and no extra monoclinic distortion is observed from 90\,K down to 40\,K. When we excite the magnetite crystal with the 400\,nm (3.10\,eV) light, we trigger multiple electronic excitations (see SI). The most dominant excitation is the charge transfer from 2p bands of oxygen to 3d bands of Fe$_{\mathrm{B}}$ \cite{Park1998}. By activating the ligand-metal charge transfer, the oxygens of the octahedron FeO$_6$ supply electrons to the Fe$_\mathrm{B}^{3+}$ non-participating ions, which consequently become Fe$_\mathrm{B}^{2+}$. Hence, the proportion of Fe$_\mathrm{B}^{2+}$ increases at the inactive trimerons B-sites, boosting t$_{2g}$ orbital ordering and creating extra trimerons, thus pushing beyond the limit of the thermodynamic maximum connectivity state leading to a new light-induced phase. This causes additional stress on the magnetite structure provoked by cooperative effects. One is the Jahn-Teller distortion due to the extending t$_{2g}$ orbital ordering. At the same time, the second contribution comes from the shortening of the atomic Fe$_\mathrm{B}$ - Fe$_\mathrm{B}$\ distances induced by the charge localization within the newly formed trimerons (charge ordering) \cite{Senn2012,Senn2015}. Recently, a high-accuracy x-ray experiment has shown that chemically doped magnetite introduces B site-selective Fe$^{2+}$ vacancies (which become Fe$^{3+}$) and weakens the trimeron long range order \cite{Pachoud2020}. The authors attribute the vanishing of the VT in doped magnetite to the absence of the trimeron network. In our case, the 400\,nm photo-doping acts oppositely and provides additional electrons to the incomplete trimeron network. As our data clearly shows, this results in an expansion of the lattice along the [110] direction. It is reasonable to speculate that the photo-doping is site-selective, which orders and strengthens the trimeron crystal seen in the increase of the coherence length (Fig.\ref{fig:2}(f)). In this scenario, the trimeron network can be seen as an imperfect Wigner crystal with homogeneous vacancies (inactive Fe$^{3+}$ B-sites). After 400\,nm photo-doping, the electrons from the oxygens activate the missing sites and complete the trimeron crystal enhancing the connectivity, which ultimately induces a monoclinic distortion favored by a strong electron-phonon coupling. 

\section{\label{sec:level5}Conclusion}
Our results reveal the emergence of two distinct metastable hidden phases in magnetite. Starting from the same LT equilibrium state, we drive magnetite into two different structural states using two different photon energies. The 800\,nm light induces d-d excitations, which favor the destruction of the trimerons and their network in a percolating fashion, leading to a phase separation between monoclinic-insulating regions and cubic-metallic islands. The 400\,nm light triggers charge transfer between oxygen and iron at B-sites, which are found to be prolific for the trimerons and their arrangement, enhancing their connectivity, which leads to a stronger monoclinic distortion inaccessible adiabatically. Our findings demonstrate the key role of the trimeron structural configuration in magnetite and show the ability to establish novel hidden phases in quantum materials via specific electronic excitations in a strongly correlated environment. 
\begin{acknowledgments}
The authors are grateful to J. Lorenzana, W. Tabis, A. Kozlowski and P. Piekarz for constructive discussions. This work was supported by the ERC consolidator grant ISCQuM No. 771346 , SNSF grant No.514725, NCCR-MUST No. 565194.
\end{acknowledgments}

\bibliographystyle{unsrt} 
\bibliography{magnetiteprl}
\end{document}


\preprint{APS/123-QED}

\title{Supplemental Material: Ultrafast generation of hidden phases \\ via energy-tuned electronic photoexcitation in magnetite}
\author{B. Truc}
\affiliation{Institute of Physics, LUMES, École Polytechnique Fédérale de Lausanne (EPFL), Lausanne CH‐1015, Switzerland}
\author{P. Usai}
\affiliation{Institute of Physics, LUMES, École Polytechnique Fédérale de Lausanne (EPFL), Lausanne CH‐1015, Switzerland}
\author{F. Pennacchio}
\affiliation{Institute of Physics, LUMES, École Polytechnique Fédérale de Lausanne (EPFL), Lausanne CH‐1015, Switzerland}
\author{G. Berruto}
\affiliation{Institute of Physics, LUMES, École Polytechnique Fédérale de Lausanne (EPFL), Lausanne CH‐1015, Switzerland}
\author{R.Claude}
\affiliation{Institute of Physics, LUMES, École Polytechnique Fédérale de Lausanne (EPFL), Lausanne CH‐1015, Switzerland}
\author{I. Madan}
\affiliation{Institute of Physics, LUMES, École Polytechnique Fédérale de Lausanne (EPFL), Lausanne CH‐1015, Switzerland}
\author{V. Sala}
\affiliation{Dipartimento di Fisica, Politecnico di Milano, Piazza Leonardo da Vinci 32, Milano, Italy}
\author{T. LaGrange}
\affiliation{Institute of Physics, LUMES, École Polytechnique Fédérale de Lausanne (EPFL), Lausanne CH‐1015, Switzerland}
\author{G. M. Vanacore}
\altaffiliation[Currently at ]{Department of Materials Science, LUMiNaD, University of Milano-Bicocca, Via Cozzi 55, 20125 Milan, Italy}
\affiliation{Institute of Physics, LUMES, École Polytechnique Fédérale de Lausanne (EPFL), Lausanne CH‐1015, Switzerland}
\author{S. Benhabib}
\email{siham.benhabib@epfl.ch}
\affiliation{Institute of Physics, LUMES, École Polytechnique Fédérale de Lausanne (EPFL), Lausanne CH‐1015, Switzerland}
\author{F. Carbone}
\email{fabrizio.carbone@epfl.ch}
\affiliation{Institute of Physics, LUMES, École Polytechnique Fédérale de Lausanne (EPFL), Lausanne CH‐1015, Switzerland}

\date{\today}

\maketitle

\section{\label{sec:level1}Experimental method }
\begin{figure} 
\begin{center}
\includegraphics[width=0.8\columnwidth,trim={2cm 1cm 2cm 2cm},clip]{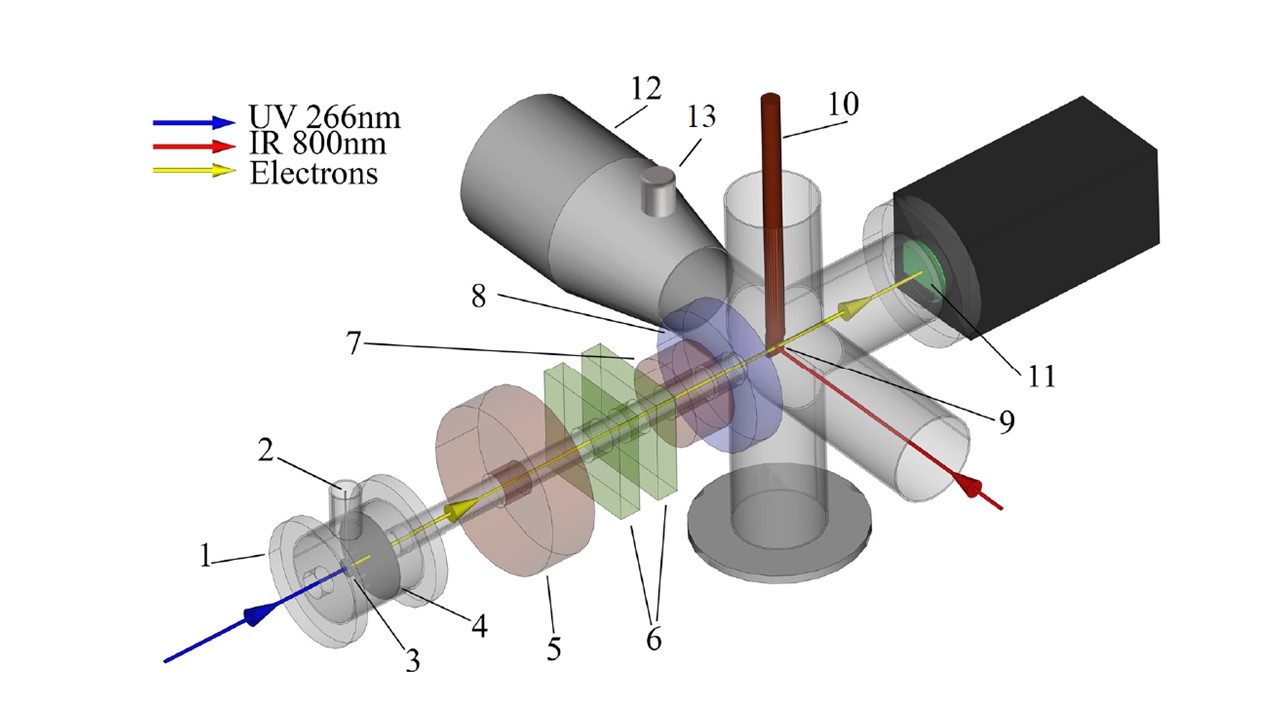}
\caption{Sketch of the UED setup at EPFL, 1) Electron gun, 2) HV connector, 3) Photocathode, 4) Anode, 5) Collimating solenoid, 6) Steering plates, 7) Focusing solenoid, 8) RF cavity, 9) Sample holder, 10) Cryostat, 11) Electron detector, 12) Ion pump, 13) Ion gauge }
\label{fig_SI:1}
\end{center}
\end{figure}

The diffraction patterns are obtained using time-resolved high-energy electron diffraction in a reflection geometry (tr-RHEED, Fig.\ref{fig_SI:1}), with a grazing angle of 0.5$^{\circ}$ to 5$^{\circ}$. The light source is a Ti:sapphire laser with a central wavelength of 800\,nm with a pulse duration of 45\,fs at a repetition rate of 20\,kHz. The beam is split into two optical paths, one for the pump (light-excitation) and the second for the probe. To generate the electrons, the third harmonic generation process is used to obtain UV of a wavelength of 266\,nm, which illuminates a silver-coated sapphire photocathode. The emitted electrons, around 10$^{5}$-10$^{6}$ electrons per pulse, are accelerated with a high voltage of 30\,keV. To overcome the space-charge effect \cite{MANCINI2012}, a 3\,GHz radiofrequency cavity temporally compresses the electron beam. The setup is combined with a designed optimal laser front tilting scheme to correct the velocity mismatch between the optical pump and the electrons at the sample surface, allowing an overall temporal resolution of $\approx$ 500\,fs \cite{Pennacchio2017}. Data were collected using two different cameras, PI-MAX 1300 Princeton Charge-Coupled Device with a number of pixels of 1340$\times$1300, with a pixel size of 20$\times$20 $\mu$m$^{2}$ combined with a microchannel plate and a direct electron (QUADRO-DECTRIS) with a resolution of 514$\times$514 with a pixel size of 75$\times$75 $\mu$m$^{2}$. The sample was fixed on a cold finger attached to a five-axis manipulator with silver conductive paste and placed inside an ultra-high vacuum chamber ($\leqslant$10$^{-9}$ mbar). The temperature range is controlled from 300\,K down to 10\,K using an open cycle cryostat Advanced Research Systems (ARS) with helium liquid flow.

\section{\label{sec:level2}Sample characterization  }
\begin{figure}[h!]
\begin{center} 
\includegraphics[width=0.8\columnwidth,trim={2cm 2cm 2cm 2cm},clip]{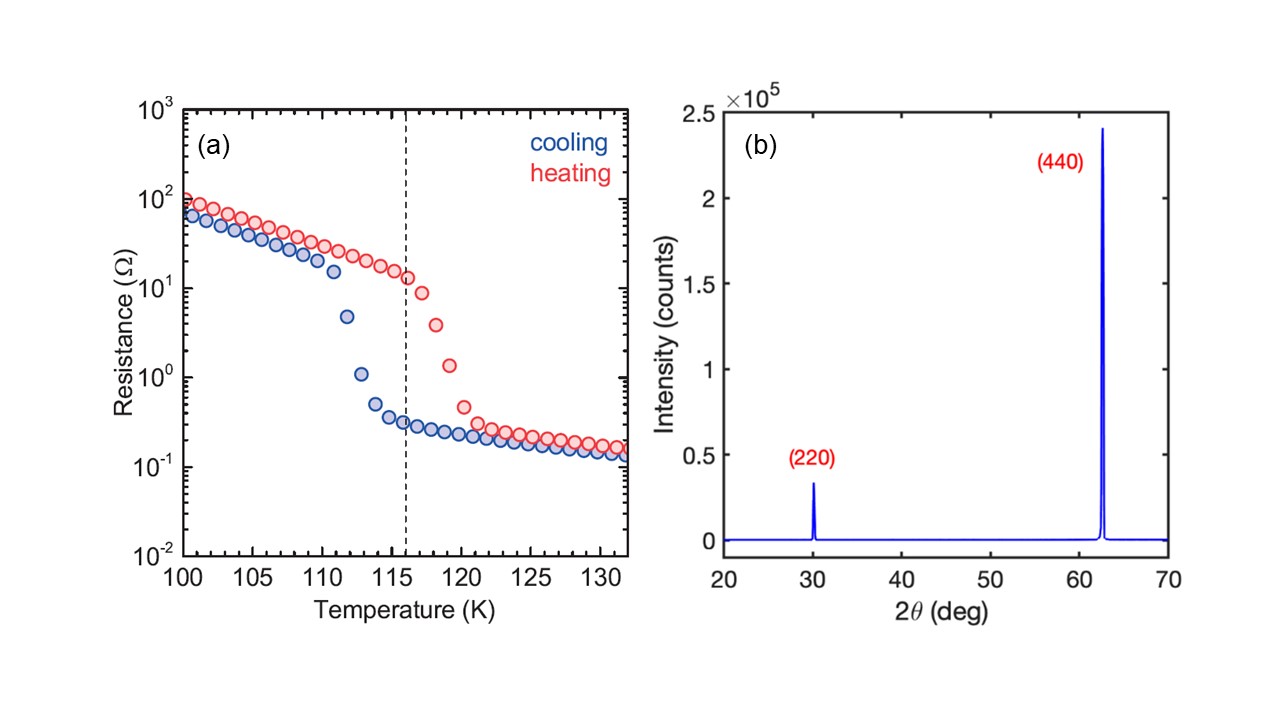}
\caption{(a): Temperature dependence of magnetite resistance. (b): Diffractogram of the single crystal Fe$_{3}$O$_{4}$ used in the experiment exposing the (110) surface, two sharp peaks identified as the (220) and the (440) planes give an average lattice parameter of a $=$ 8.3853 $\AA$ at room temperature.}
\label{fig_SI:2} 
\end{center}
\end{figure}

The magnetite sample is a single crystal. It was characterized by transport measurements (see Fig.\ref{fig_SI:2} (a)). The resistance shows a first-order transition at a Verwey temperature  T$_{V}\approx$117\,K. The sample surface with an out-of-plane direction of [110] is polished to get optical flatness. The orientation was checked by x-ray diffraction with an Empyrean diffractometer from PANanalytical, using Cu-K$\alpha$1 and Cu-K$\alpha$2 radiations at ambient conditions (see Fig.\ref{fig_SI:2} (b)).

\section{\label{sec:level3}The atomic displacement and strain }
\begin{figure}[h!]
\begin{center}
\includegraphics[width=0.8\columnwidth,trim={4cm 2cm 4cm 2cm},clip]{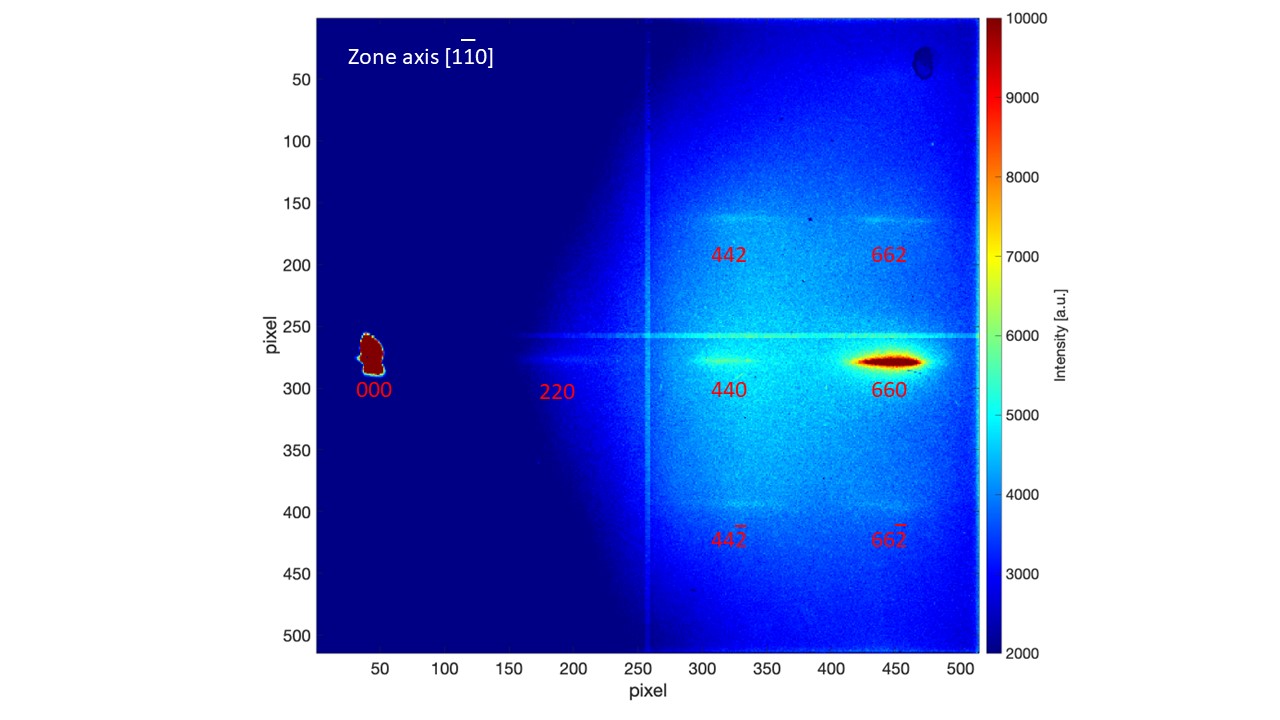}
\caption{Assignment of the Bragg peaks visible on our camera along the [1$\bar{1}$0] zone axis with 30 keV electron beam. }
\label{fig_SI:3}
\end{center}
\end{figure}

The atomic displacement (compression or expansion) was extracted from the Bragg peak position with the following method:
\begin{enumerate}
 \item The atomic displacement ($u$) is linked to the initial atomic interplanar distance $d'_{hkl}$ and the final atomic interplanar distance $d_{hkl}$ by:
 \[u=\frac{d'_{hkl}-d_{hkl}}{d_{hkl}}\] 
 
  \item The Bragg peak position corresponds to a specific diffraction angle $\theta$, which is related to the interatomic distance by the Bragg law:
  \[d_{hkl}=\frac{n\lambda}{2\sin\theta}\] 
   $\lambda=0.069\mathrm{\AA}$ is the electron wavelength at 30\,kV. 
   \item In our case $\theta$ varies from 0.5$^{\circ}$ to 5$^{\circ}$, thus, in  the small-angle approximation, the atomic displacement is given by:
   \[u=\frac{\frac{1}{\theta'}-\frac{1}{\theta}}{\frac{1}{\theta}}=\frac{\theta-\theta'}{\theta'}\] 
\end{enumerate}
In this study, the electron beam goes along the [1$\bar{1}$0] zone axis of the crystal probing the atomic displacement along the [110] direction. The different Bragg peaks visible in that configuration are presented in Fig.\ref{fig_SI:3}. Note that the (660) peak is the most intense.\\

In the theory of elasticity, the deformation is a pure strain, which is defined  by the atomic displacement ($u$) and the atomic position ($x$) \cite{Luthi2005}:

\[\varepsilon_{ij}=\frac{1}{2}\left(\frac{\partial u_{i}}{\partial x_{j}}+\frac{\partial u_{j}}{\partial x_{i}}\right)\]

The presence of strain ($\varepsilon$) induces a stress ($\sigma$). These two quantities are related by the elastic tensor $c_{ijkl}$ through Hooke’s law:    
\[\sigma_{ij}=\sum_{kl}c_{ijkl}\varepsilon_{kl}\].
For the cubic lattice, the elastic tensor $c_{ijkl}$ is given by:
\begin{equation*}
  \begin{pmatrix}
         \sigma_{xx} \\
         \sigma_{yy} \\ 
         \sigma_{zz} \\
         \sigma_{yz} \\ 
         \sigma_{xz} \\ 
        \sigma_{xy}
     \end{pmatrix}
     =
     \begin{pmatrix}
         c_{11} & c_{12} & c_{12} & 0 & 0 & 0\\
         c_{12} & c_{12} & c_{12} & 0 & 0 & 0\\ 
         c_{12} & c_{12} & c_{11} & 0 & 0 & 0\\
         0 & 0 & 0 & c_{44} & 0 & 0\\
         0 & 0 & 0 & 0 &c_{44} & 0\\
         0 & 0 & 0 & 0 & 0 & c_{44}\\
     \end{pmatrix}
     \times
     \begin{pmatrix}
         \varepsilon_{xx} \\
         \varepsilon_{yy} \\ 
         \varepsilon_{zz} \\
        2\varepsilon_{yz} \\ 
        2\varepsilon_{xz} \\ 
        2\varepsilon_{xy}
     \end{pmatrix}
\end{equation*}
Therefore, the deformation probed along [110] corresponds to $\varepsilon_{xy}$ and the related stress $\sigma_{xy}$ is:
\[\sigma_{xy}=2c_{44}\varepsilon_{xy}\]
In the case of the monoclinic structure, the elastic tensor $c_{ijkl}$ is: 
\begin{equation*}
  \begin{pmatrix}
         \sigma_{xx} \\
         \sigma_{yy} \\ 
         \sigma_{zz} \\
         \sigma_{yz} \\ 
         \sigma_{xz} \\ 
        \sigma_{xy}
     \end{pmatrix}
     =
     \begin{pmatrix}
         c_{11} & c_{12} & c_{13} & 0 & c_{15} & 0\\
         c_{12} & c_{22} & c_{23} & 0 & c_{25} & 0\\ 
         c_{13} & c_{23} & c_{33} & 0 & c_{35} & 0\\
         0 & 0 & 0 & c_{44} & 0 & c_{46}\\
         c_{15} & c_{25} & c_{35} & 0 & c_{55} & 0\\
         0 & 0 & 0 &c_{46} & 0 & c_{66}\\
     \end{pmatrix}
     \times
     \begin{pmatrix}
         \varepsilon_{xx} \\
         \varepsilon_{yy} \\ 
         \varepsilon_{zz} \\
         2\varepsilon_{yz} \\ 
        2 \varepsilon_{xz} \\ 
        2\varepsilon_{xy}
     \end{pmatrix}
\end{equation*}
Therefore, the deformation probed along [110] corresponds to $\varepsilon_{xy}$ and the related stress $\sigma_{xy}$ is:
\[\sigma_{xy}=2(c_{46}\varepsilon_{yz}+c_{66}\varepsilon_{xy})\]

\section{\label{sec:level4}Ginzburg-Landau analysis (Symmetry)}
It is instructive to retrieve the symmetry of the order parameter (OP), based on the framework of Ginzburg-Landau’s theory \cite{Landau1980} and a fundamental group theory analysis. Briefly, the dominating term in the free energy $F$, is the linear coupling between the strain and the OP expressed as:
\[F_{coupling}= g\varepsilon \Delta\],
where $g$ is the coupling constant, $\varepsilon $ is the strain, and $\Delta$ is the order parameter. The free energy coupling term is part of the total free energy and must remain invariant under all symmetry operations. In the high-temperature phase, the point group of magnetite is $O_{h}$. Consequently, the only representation in the character table (see Table.\ref{tab:table1}) respecting this constrain is the $A_{1g}$ representation. Hence, the symmetry product between $\varepsilon_{xy}$ and $\Delta$ should belong to the $A_{1g}$ representation (see Table.\ref{tab:table2}). Since the shear strain $\varepsilon_{xy}$ belongs to the $T_{2g}$ representation, the order parameter $\Delta$ must belong to the $T_{2g}$ representation as well.
\begin{table*}[h!]
\begin{center}
\begin{tabular}{|c|cccccccccc|c|c|}
\hline
\boldmath$O_h$ &  &  &  &  &  &  &  &  &  &  & linear functions, & quadratic functions \\ 
group & \boldmath$E$ & \boldmath$8C_3$ & \boldmath$6C_2$ & \boldmath$6C_4$ & \boldmath$3C_2$ & \boldmath$i$ & \boldmath$6S_4$ & \boldmath$8S_6$ & \boldmath$3\sigma_h$ & \boldmath$6\sigma_d$ & rotations & \\
\hline
\textbf{A}\boldmath$_{1\mathrm{g}}$ & +1 & +1 & +1 & +1 & +1 & +1 & +1 & +1 & +1 & +1 & - & $x^2+y^2+z^2$ \\
\textbf{A}\boldmath$_{2\mathrm{g}}$ & +1 & +1 & -1 & -1 & +1 & +1 & -1 & +1 & +1 & -1 & - & - \\
\textbf{E}\boldmath$_{\mathrm{g}}$ & +2 & -1 & 0 & 0 & +2 & +2 & 0 & -1 & +2 & 0 & - & $(2z^2-x^2-y^2, x^2-y^2)$ \\
\textbf{T}\boldmath$_{1\mathrm{g}}$ & +3 & 0 & -1 & +1 & -1 & +3 & -1 & 0 & -1 & -1 & $(R_x, R_y, R_z)$ & - \\
\textbf{T}\boldmath$_{2\mathrm{g}}$ & +3 & 0 & +1 & -1 & -1 & +3 & -1 & 0 & -1 & +1 & - & $(xz,yz,xy)$ \\
\textbf{A}\boldmath$_{1\mathrm{u}}$ & +1 & +1 & +1 & +1 & +1 & -1 & -1 & -1 & -1 & -1 & - & - \\
\textbf{A}\boldmath$_{2\mathrm{u}}$ & +1 & +1 & -1 & -1 & +1 & -1 & +1 & -1 & -1 & +1 & - & - \\
\textbf{E}\boldmath$_{\mathrm{u}}$ & +2 & -1 & 0 & 0 & +2 & -2 & 0 & +1 & -2 & 0 & - & - \\
\textbf{T}\boldmath$_{1\mathrm{u}}$ & +3 & 0 & -1 & +1 & -1 & -3 & -1 & 0 & +1 & +1 & $(x,y,z)$ & - \\
\textbf{T}\boldmath$_{2\mathrm{u}}$ & +3 & 0 & +1 & -1 & -1 & -3 & +1 & 0 & +1 & -1 & - & - \\
\hline 
\end{tabular}
\caption{\label{tab:table1}Characters table for $O_h$ point group.}
\end{center}
\end{table*}

\begin{figure}[h!]
\begin{center}
\includegraphics[width=0.9\columnwidth,trim={2cm 5cm 2cm 2cm},clip]{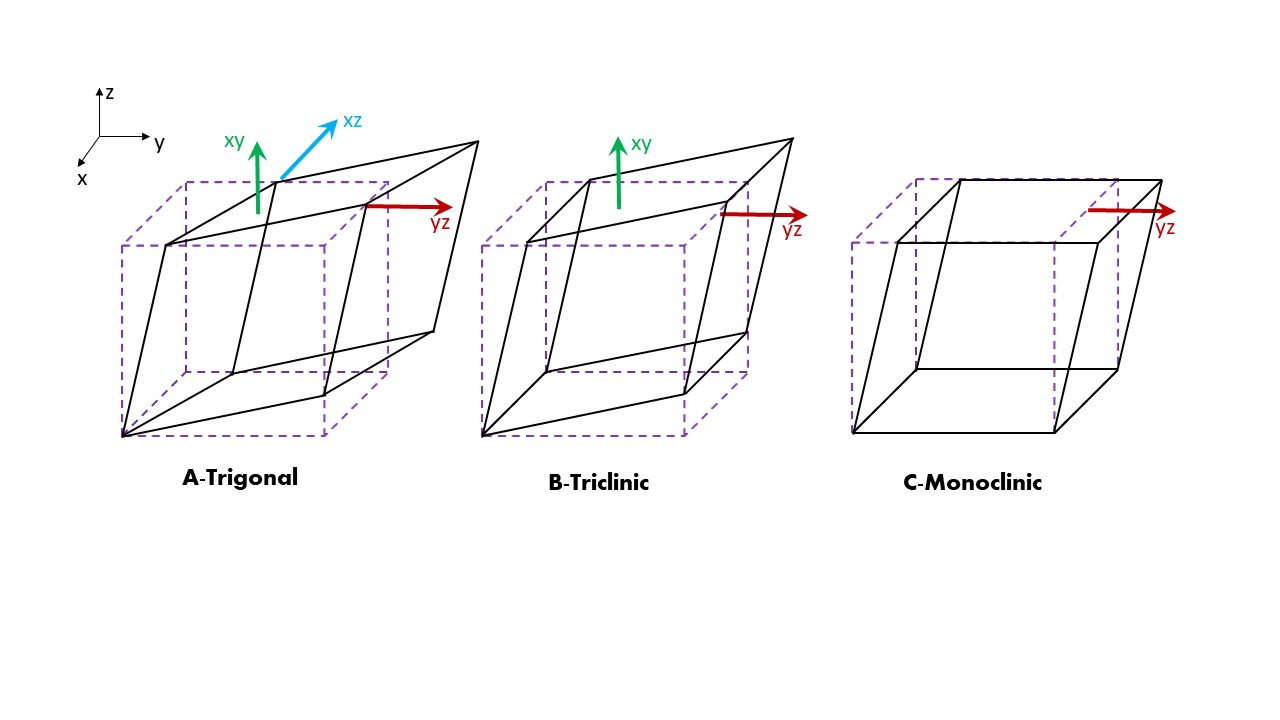}
\caption{The sketch of the three possible structural transformations from the cubic structure resulting from T$_{2g}$ symmetry. The dashed purple structure corresponds to the initial cubic structure. The arrows represent the strain directions (red =$\epsilon_{yz}$,  green=$\epsilon_{xy}$, blue=$\epsilon_{xz}$)}
\label{fig_SI:4}
\end{center}
\end{figure}

\begin{center}
\begin{table*}[h!]
\scriptsize
\hspace{-1.5cm}
\begin{tabular}{|c|c|c|c|c|c|c|c|c|c|c|}
\hline 
& \textbf{A}\boldmath$_{1\mathrm{g}}$ & \textbf{A}\boldmath$_{2\mathrm{g}}$ & \textbf{E}\boldmath$_{\mathrm{g}}$ & \textbf{T}\boldmath$_{1\mathrm{g}}$ & \textbf{T}\boldmath$_{2\mathrm{g}}$ & \textbf{A}\boldmath$_{1\mathrm{u}}$ & \textbf{A}\boldmath$_{2\mathrm{u}}$ & \textbf{E}\boldmath$_{\mathrm{u}}$ & \textbf{T}\boldmath$_{1\mathrm{u}}$ & \textbf{T}\boldmath$_{2\mathrm{u}}$ \\ 
\hline 
\textbf{A}\boldmath$_{1\mathrm{g}}$ & A$_{1\mathrm{g}}$ & A$_{2\mathrm{g}}$ & E$_{\mathrm{g}}$ & T$_{1\mathrm{g}}$ & T$_{2\mathrm{g}}$ & A$_{1\mathrm{u}}$ & A$_{2\mathrm{u}}$ & E$_{\mathrm{u}}$ & T$_{1\mathrm{u}}$ & T$_{2\mathrm{u}}$ \\ \hline
\textbf{A}\boldmath$_{2\mathrm{g}}$ & A$_{2\mathrm{g}}$ & A$_{1\mathrm{g}}$ & E$_{\mathrm{g}}$ & T$_{2\mathrm{g}}$ & T$_{1\mathrm{g}}$ & A$_{2\mathrm{u}}$ & A$_{1\mathrm{u}}$ & E$_{\mathrm{u}}$ & T$_{2\mathrm{u}}$ & T$_{1\mathrm{u}}$ \\ \hline
\textbf{E}\boldmath$_{\mathrm{g}}$ & E$_{\mathrm{g}}$ & E$_{\mathrm{g}}$ & A$_{1\mathrm{g}}$+A$_{2\mathrm{g}}$+E$_{\mathrm{g}}$& T$_{1\mathrm{g}}$+T$_{2\mathrm{g}}$ & T$_{1\mathrm{g}}$+T$_{2\mathrm{g}}$ & E$_{\mathrm{u}}$ & E$_{\mathrm{u}}$ & A$_{1\mathrm{u}}$+A$_{2\mathrm{u}}$+E$_{\mathrm{u}}$ & T$_{1\mathrm{u}}$+T$_{2\mathrm{u}}$ & T$_{1\mathrm{u}}$+T$_{2\mathrm{u}}$ \\ \hline 
\textbf{T}\boldmath$_{1\mathrm{g}}$ & T$_{1\mathrm{g}}$ & T$_{2\mathrm{g}}$ & T$_{1\mathrm{g}}$+T$_{2\mathrm{g}}$ & A$_{1\mathrm{g}}$+E$_{\mathrm{g}}$+T$_{1\mathrm{g}}$+T$_{2\mathrm{g}}$ & A$_{2\mathrm{g}}$+E$_{\mathrm{g}}$+T$_{1\mathrm{g}}$+T$_{2\mathrm{g}}$ & T$_{1\mathrm{u}}$ & T$_{2\mathrm{u}}$ & T$_{1\mathrm{u}}$+T$_{2\mathrm{u}}$& A$_{1\mathrm{u}}$+E$_{\mathrm{u}}$+T$_{1\mathrm{u}}$+T$_{2\mathrm{u}}$ & A$_{2\mathrm{u}}$+E$_{\mathrm{u}}$+T$_{1\mathrm{u}}$+T$_{2\mathrm{u}}$ \\ \hline  
\textbf{T}\boldmath$_{2\mathrm{g}}$ & T$_{2\mathrm{g}}$ & T$_{1\mathrm{g}}$ & T$_{1\mathrm{g}}$+T$_{2\mathrm{g}}$ & A$_{2\mathrm{g}}$+E$_{\mathrm{g}}$+T$_{1\mathrm{g}}$+T$_{2\mathrm{g}}$ & A$_{1\mathrm{g}}$+E$_{\mathrm{g}}$+T$_{1\mathrm{g}}$+T$_{2\mathrm{g}}$ & T$_{2\mathrm{u}}$ & T$_{1\mathrm{u}}$ & T$_{1\mathrm{u}}$+T$_{2\mathrm{u}}$& A$_{2\mathrm{u}}$+E$_{\mathrm{u}}$+T$_{1\mathrm{u}}$+T$_{2\mathrm{u}}$ & A$_{1\mathrm{u}}$+E$_{\mathrm{u}}$+T$_{1\mathrm{u}}$+T$_{2\mathrm{u}}$ \\ \hline 
\textbf{A}\boldmath$_{1\mathrm{u}}$ & A$_{1\mathrm{u}}$ & A$_{2\mathrm{u}}$ & E$_{\mathrm{u}}$ & T$_{1\mathrm{u}}$ & T$_{2\mathrm{u}}$ & A$_{1\mathrm{g}}$ & A$_{2\mathrm{g}}$ & E$_{\mathrm{g}}$ & T$_{1\mathrm{g}}$ & T$_{2\mathrm{g}}$ \\ \hline  
\textbf{A}\boldmath$_{2\mathrm{u}}$ & A$_{2\mathrm{u}}$ & A$_{1\mathrm{u}}$ & E$_{\mathrm{u}}$ & T$_{2\mathrm{u}}$ & T$_{1\mathrm{u}}$ & A$_{2\mathrm{g}}$ & A$_{1\mathrm{g}}$ & E$_{\mathrm{g}}$ & T$_{2\mathrm{g}}$ & T$_{1\mathrm{g}}$ \\ \hline 
\textbf{E}\boldmath$_{\mathrm{u}}$ & E$_{\mathrm{u}}$ & E$_{\mathrm{u}}$ & A$_{1\mathrm{u}}$+A$_{2\mathrm{u}}$+E$_{\mathrm{u}}$& T$_{1\mathrm{u}}$+T$_{2\mathrm{u}}$ & T$_{1\mathrm{u}}$+T$_{2\mathrm{u}}$ & E$_{\mathrm{g}}$ & E$_{\mathrm{g}}$ & A$_{1\mathrm{g}}$+A$_{2\mathrm{g}}$+E$_{\mathrm{g}}$ & T$_{1\mathrm{g}}$+T$_{2\mathrm{g}}$ & T$_{1\mathrm{g}}$+T$_{2\mathrm{g}}$ \\ \hline 
\textbf{T}\boldmath$_{1\mathrm{u}}$ & T$_{1\mathrm{u}}$ & T$_{2\mathrm{u}}$ & T$_{1\mathrm{u}}$+T$_{2\mathrm{u}}$ & A$_{1\mathrm{u}}$+E$_{\mathrm{u}}$+T$_{1\mathrm{u}}$+T$_{2\mathrm{u}}$ & A$_{2\mathrm{u}}$+E$_{\mathrm{u}}$+T$_{1\mathrm{u}}$+T$_{2\mathrm{u}}$ & T$_{1\mathrm{g}}$ & T$_{2\mathrm{g}}$ & T$_{1\mathrm{g}}$+T$_{2\mathrm{g}}$& A$_{1\mathrm{g}}$+E$_{\mathrm{g}}$+T$_{1\mathrm{g}}$+T$_{2\mathrm{g}}$ & A$_{2\mathrm{g}}$+E$_{\mathrm{g}}$+T$_{1\mathrm{g}}$+T$_{2\mathrm{g}}$ \\ \hline 
\textbf{T}\boldmath$_{2\mathrm{u}}$ & T$_{2\mathrm{u}}$ & T$_{1\mathrm{u}}$ & T$_{1\mathrm{u}}$+T$_{2\mathrm{u}}$ & A$_{2\mathrm{u}}$+E$_{\mathrm{u}}$+T$_{1\mathrm{u}}$+T$_{2\mathrm{u}}$ & A$_{1\mathrm{u}}$+E$_{\mathrm{u}}$+T$_{1\mathrm{u}}$+T$_{2\mathrm{u}}$ & T$_{2\mathrm{g}}$ & T$_{1\mathrm{g}}$ & T$_{1\mathrm{g}}$+T$_{2\mathrm{g}}$& A$_{2\mathrm{g}}$+E$_{\mathrm{g}}$+T$_{1\mathrm{g}}$+T$_{2\mathrm{g}}$ & A$_{1\mathrm{g}}$+E$_{\mathrm{g}}$+T$_{1\mathrm{g}}$+T$_{2\mathrm{g}}$ \\
\hline 
\end{tabular}
\caption{Product table of the $O_h$ point group.}
\label{tab:table2}
\end{table*}
\end{center}
The $T_{2g}$  symmetry implies that the OP contains three components:
\[\Delta=(\Delta_{xy}+\Delta_{yz}+\Delta_{xz})\]
Thus, we can rewrite the coupling free energy term as:
\[F_{coupling}= g(\varepsilon_{xy} \Delta_{xy}+\varepsilon_{yz} \Delta_{yz}+\varepsilon_{xz} \Delta_{xz})\]
Three possible scenarios follow (See Fig.\ref{fig_SI:4}):
\begin{enumerate}
\item $\Delta_{xy}, \Delta_{yz}, \Delta_{xz}\Longrightarrow  \varepsilon_{xy}, \varepsilon_{yz}, \varepsilon_{xz}\neq 0$\\
Applying simultaneously three shear strains  $\varepsilon_{xy}, \varepsilon_{yz}, \varepsilon_{xz}$ leads to a deviation of the angles from 90$^{\circ}$, without any changes in the lattice parameters, a, b, and c yielding a trigonal distortion (case A in Fig.\ref{fig_SI:4}).
\item $\Delta_{xy}, \Delta_{yz}\neq 0 $ and $ \Delta_{xz}=0 \Longrightarrow \varepsilon_{xy},\varepsilon_{yz}\neq 0$ and $ \varepsilon_{xz}=0 $\\
Applying simultaneously two shear strains $\varepsilon_{xy}$  and $\varepsilon_{yz}$  induces the derivation of the angles from the 90$^{\circ}$ accompanied by nonequivalent lattice parameters $a\neq b \neq c$ generating a triclinic distortion (case B in Fig.\ref{fig_SI:4}).
\item $\Delta_{yz}, \Delta_{xz}= 0 $ and $ \Delta_{xy}\neq 0 \Longrightarrow \varepsilon_{yz}, \varepsilon_{xz}= 0 $ and $ \varepsilon_{xy}\neq 0$\\
Applying only one shear strain $\varepsilon_{xy}$ modifies the lattice parameter c and keeps a and b equivalent, and one angle is different from 90$^{\circ}$, leading to a monoclinic distortion (case C in Fig.\ref{fig_SI:4}). 
\end{enumerate}
The latter case agrees with the low-temperature structure of magnetite $Cc$ \cite{Iizumi1982}. Hence, we conclude that the OP responsible for the structural change belongs to $T_{2g}$ with one component $\Delta=(\Delta_{xy},0,0)$.

\section{\label{sec:level5}Estimation of the temperature rise }
We estimate a maximum temperature rise due to energy deposed by the laser pulses with two different models, a simple heating model, and the Debye-Waller model. 
\begin{enumerate}
    \item \textbf{Estimation based on the heating model}\\
    We estimate the temperature rise after the photo excitation with a simple heating model. For simplicity, we neglect heat-dissipation to assess the upper boundary and consider an ellipsoid box geometry that corresponds to the experimental condition.
    The model consists of solving the calorimetry equation:
  \[E_{pulse}=m\int_{T_{0}}^{T_{0}+\Delta T} C_{p}(T) \,dT \]

$E_{pulse}$ is the energy given by a single pulse, $m$ is the number of moles interacting retrieved from the volume considered, $T$ is the temperature, $C_{p}(T)$ is the molar heat capacity, taken from ref \cite{WESTRUM1969} and $T_{0}$ is the initial temperature before the photoexcitation.\\
First, we compute the total interacting volume, which is given by the product of the beam footprint and the penetration depth. The ellipsoid beam shape is characterized with a beam profiler while the penetration depth is computed from the absorption coefficient taken from ref \cite{Schlegel1979}.
This gives an estimated interaction volumes of $\sim$ 1.87\,mm $\times$ 0.22\,mm $\times$ $\pi$ $\times$ 108\,nm $=$ 1.4 10 $^{-13}$ m$^{3}$ and 1.75\,mm $\times$ 0.21\,mm $\times$ $\pi$ $\times$ 34\,mm $=$ 3.9 10$^{-14}$\,m$^{3}$ for 800\,nm and 400\,nm, respectively. 
We finally retrieved the number of moles using 231.5\,g/mol as the molar mass of Fe$_{3}$O$_{4}$ and its density of 5.17\,g/cm$^{3}$.\\
Next, we calculate the energy deposed by a single pulse $E_{pulse}$. This pulse energy is calculated as follows:
\[E_{pulse}=\frac{1-R}{e} \frac{P}{rep.rate}\]
Where $P$ is the total incident power, $rep.rate$ is the repetition rate of the laser (20\,kHz), $R$ the reflectivity, taken from ref \cite{Randi2016}. Solving numerically the calorimetric equation, we find for 800\,nm an increase in temperature of 46\,K (58\,K) for 2.9\,mJ/cm$^2$ (3.7\,mJ/cm$^2$). For 400\,nm optical excitation the increase of temperature $\Delta$T is equals to 60\,K with 1.2\,mJ/cm$^2$. Note our electron probe penetration length is $\sim$ 5\,nm one order of magnitude smaller than the optical penetration length which is $\sim$ 34\,nm for the 400\,nm.

\item \textbf{The Debye-Waller model}\\
To estimate the evolution of the lattice temperature after the photoexcitation, we used the Debye-Waller effect \cite{MO2018}. It relates how the intensity $I$ varies as a function of the temperature-dependent atomic displacement $u$.
 \[\frac{I_{hkl}}{I_{0_ {hkl}}}=exp\{-\frac{1}{3}\langle u^{2} \rangle (t) Q^{2}_{hkl}\}\]
 
 Where, $Q_{hkl}=\frac{4\pi \theta_{hkl}}{\lambda}$ is the scattering vector and $\lambda=0.069\mathrm{\AA}$ is the electron wavelength at 30\,kV.\\
 
 In Fig.\ref{fig_SI:5} (a,b), we present the evolution of the Debye-Waller factor $\langle u^{2} \rangle (t)$ after the light excitation at 800\,nm and 400\,nm, respectively, calculated from the evolution of the Bragg peak intensity using the following relation:
 \[\langle u^{2} \rangle (t)= \frac{-3\ln{\bigl(\frac{I_{hkl}}{I_{0_ {hkl}}}}\bigr)}{Q^{2}_{hkl}} \]
 
\begin{figure}[h!]
\centering 
\includegraphics[width=1\columnwidth,trim={4cm 2cm 4cm 1cm},clip]{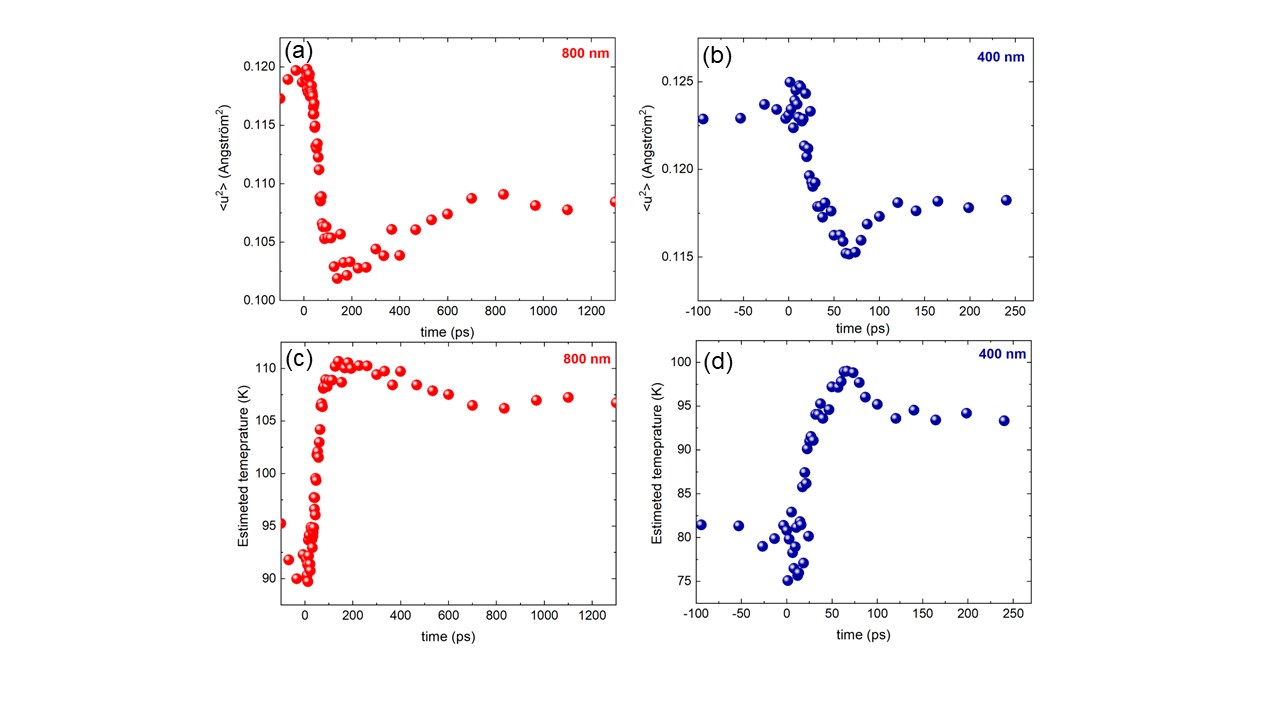}
\caption{ (a,b): the temporal evolution of the Debye-Waller factor $\langle u^{2} \rangle (t)$〉 for the 800 and 400\,nm excitations, respectively. (c,d): the temporal evolution of the estimated temperature after light excitations 800 and 400\,nm , respectively. }
\label{fig_SI:5} 
\end{figure}
 To estimate the temperature T, we use the temperature variation with $\langle u^{2} \rangle$  from the static data presented in Fig.\ref{fig_SI:6}, and map the correspondence with the relation  (fit):
 \[T(K)=71.628+50.014\bigl\{\cot(\frac{\langle u^{2} \rangle-0.126}{-0.0053}\bigr\}\]
 \begin{figure}[h!]
 \begin{center}
\includegraphics[width=0.8\columnwidth]{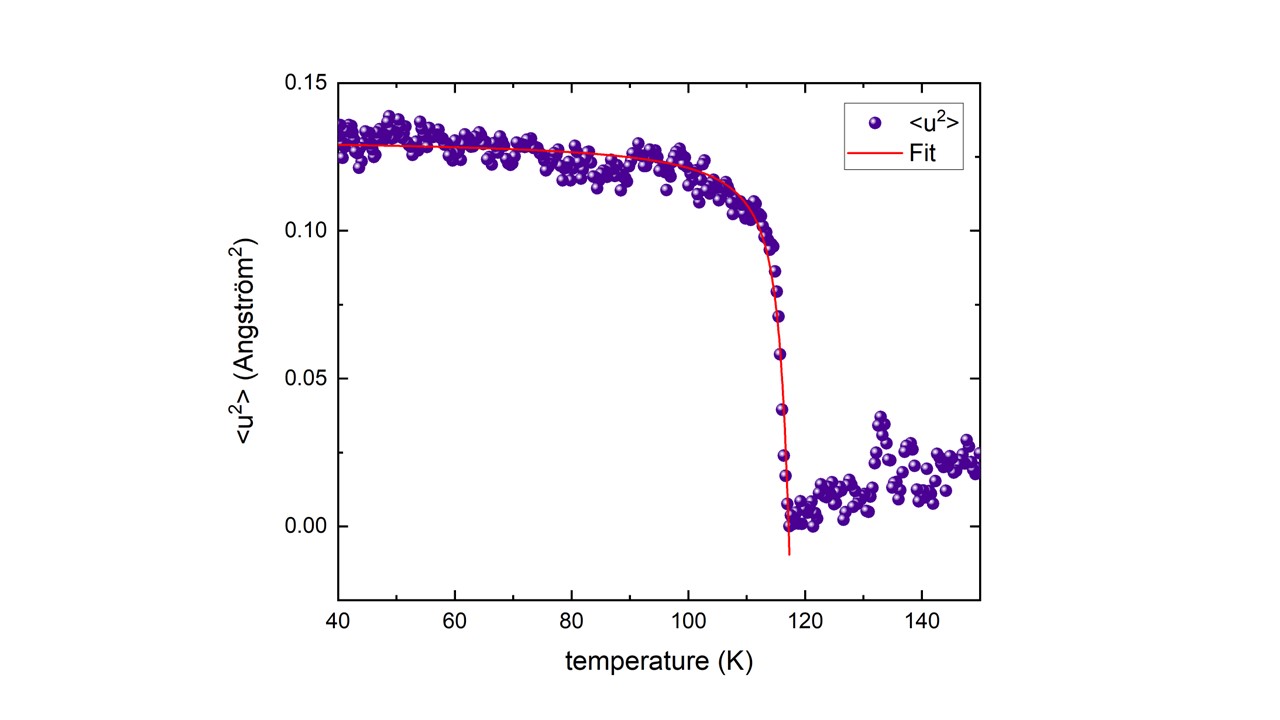}
\caption{The temperature dependence of the atomic displacement $\langle u^{2} \rangle (t)$  extracted from the temperature dependence of the (660) Bragg peak intensity. The red line is the fit. }
\label{fig_SI:6} 
 \end{center}
\end{figure}

 Finally, to get the temporal evolution of the estimated temperature (Fig.\ref{fig_SI:5}(c,d)), we convert the Debye-Waller factor to temperature using the temperature variation with $\langle u^{2} \rangle$ presented previously.\\
 We found for the 800\,nm optical excitation; the maximum temperature rise is $\Delta$ T=111-89 =22\,K. While for the 400\,nm case, the maximum temperature rise is $\Delta$T=99-80=19\,K.\\ 
The Debye-Waller gives a temperature increase of around 20\,K lower than the simple heating model. This model extracts the temperature evolution from experimental data but considers an isotropic harmonic potential and neglects other potential contributions in the intensity as the change over time of the atomic potential itself due to a structural change. Moreover, this model does not take in account the phase separation induced by the 800\,nm photoexcitation. 
\end{enumerate}

\section{\label{sec:level6}Extended delay for 800\,nm excitation}
For the 800\,nm light excitation, we extended the delay up to 1.3\,ns to follow the complete thermalization process. Although the atomic displacement seems to have almost completely relaxed, the intensity and the  full width at half maximum (FWHM) are still far from their values before the photoexcitation.\\
\begin{figure}[h!]
\begin{center}
\includegraphics[width=1\columnwidth]{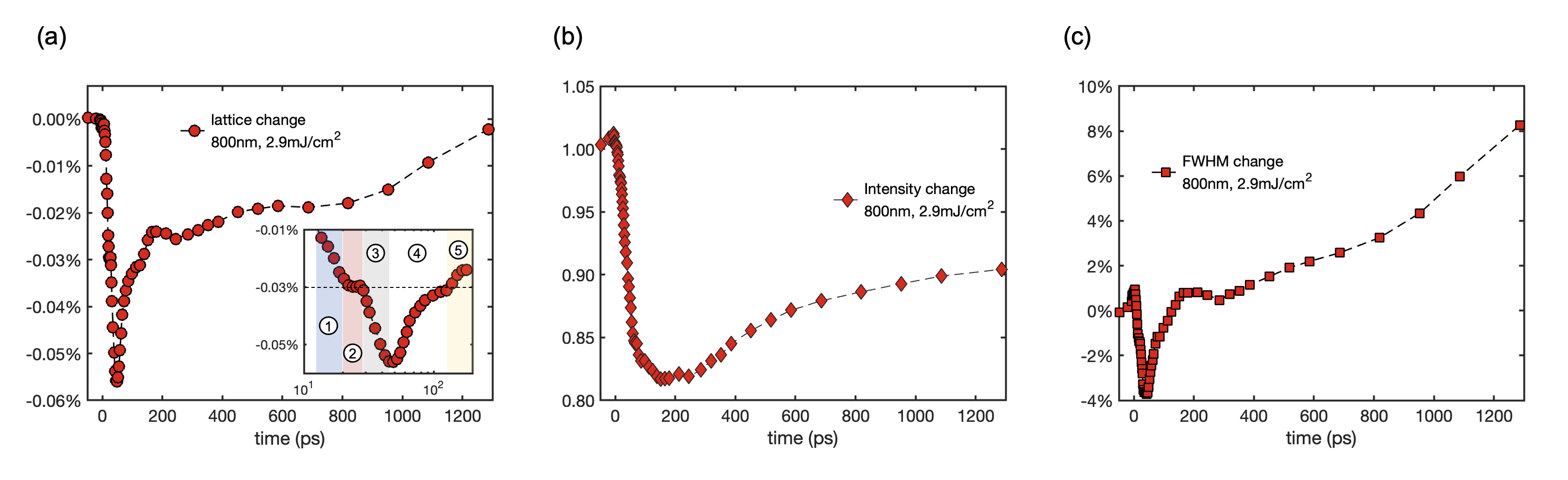}
\caption{(a), (b), (c) Evolution of the interplanar atomic displacement (Expansion/compression) along the [110] direction, normalized intensity and FWHM for the (660) Bragg peak under 800 nm photoexcitation, respectively. }
\label{fig_SI:7} 
\end{center}
\end{figure}

\section{\label{sec:level7}Second data set with 800\,nm photoexcitation }
\begin{figure}[h!]
\begin{center}
\includegraphics[width=1\columnwidth]{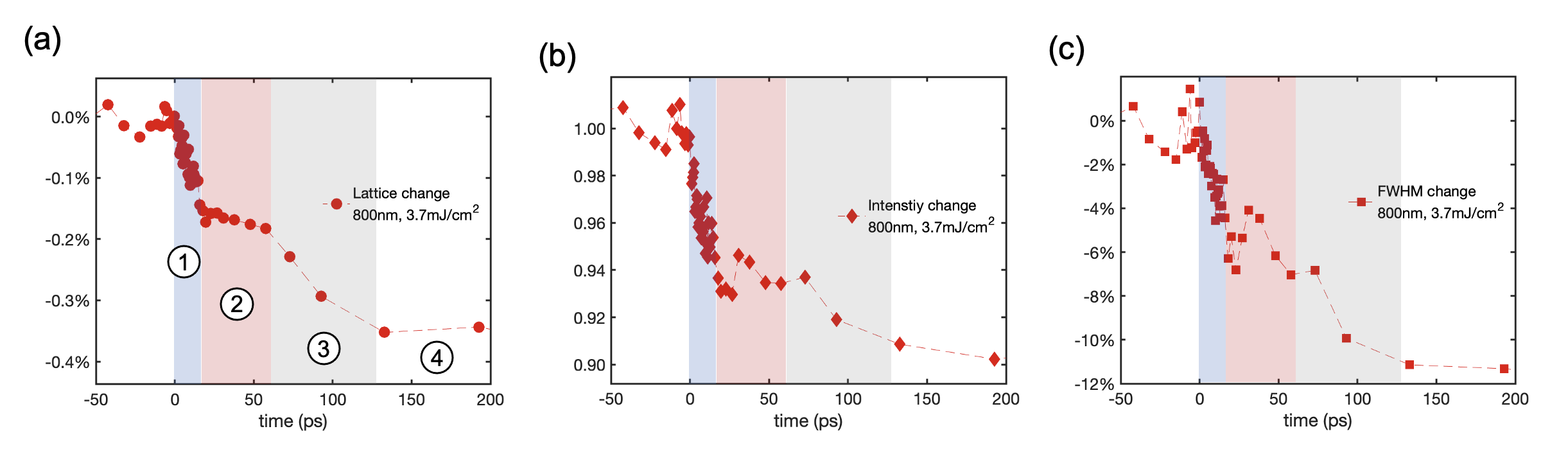}
\caption{(a), (b), (c) Evolution of the interplanar atomic displacement (Expansion/compression) along the [110] direction, normalized intensity and FWHM for the (660) Bragg peak under 800 nm photoexcitation, respectively. the shaded areas show the multiple compression stages. }
\label{fig_SI:8} 
\end{center}
\end{figure}

A second data with a higher fluence (3.7\,mJ/cm$^2$) taken with a different detector is presented in Fig.\ref{fig_SI:8}. The comparison with the data set presented in the main text demonstrates a good consistency and that the duration and the amplitude of each stage depends on the fluence used.

\section{\label{sec:level8} Other possible electronic excitations for the 400\,nm photoexcitation }
GGA+U calculation \cite{Borroni2017} predicts two other possible electronic excitations triggered by 400\,nm light: i) the transition from an occupied t$_{2g}$ of Fe$_\mathrm{B}^{2+}$ to an unoccupied e$_g$ of Fe$_\mathrm{B}^{3+}$ and ii) the transition from Fe$_\mathrm{B}^{2+}$ to Fe$_\mathrm{A}^{3+}$ with an onset of approximately 2.4\,eV and 2.7\,eV respectively. The transition from an occupied t$_{2g}$ of Fe$_\mathrm{B}^{2+}$ to an unoccupied e$_g$ of Fe$_\mathrm{B}^{3+}$ leads to valency change for both Fe$_\mathrm{B}^{2+}$ and Fe$_\mathrm{B}^{3+}$. Hence, by alternating their sites inside the trimeron unit, trimerons destruction is similar to the case of the 800\,nm excitation. However, in this case, the difference is that not only the trimeron unit is destroyed but also the t$_{2g}$ orbital ordering. Hence, we expect a compression of the lattice leading to a recovery of the cubic phase. In the second case, the transition from Fe$_\mathrm{B}^{2+}$ to Fe$_\mathrm{A}^{3+}$ leads to decreasing the number of Fe$^{2+}$ ions and increasing Fe$^{3+}$ at the B-sites, resulting in trimeron destruction and favoring the cubic phase. In both cases, we would expect a compression of the lattice along the [110] direction, which we did not observe in our data consistently because these two electronic excitation mechanisms are less probable than the dominant contribution provided by the O-Fe charge transfer.
\section{\label{sec:level9} controlling the MIT by ultrafast light pulses in magnetite }
\begin{figure}[h!]
\begin{center}
\includegraphics[width=0.8\columnwidth]{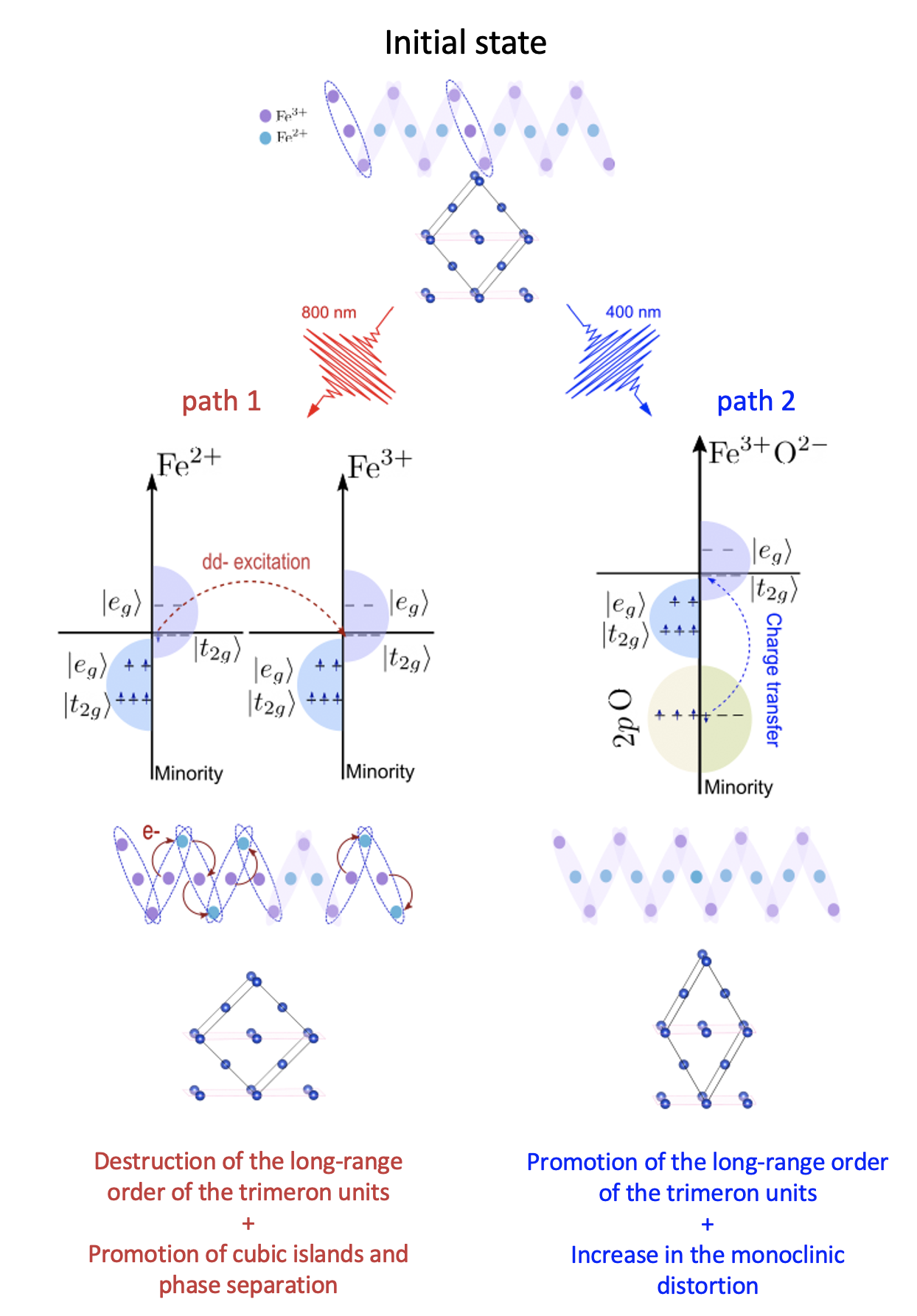}
\caption{A demonstrative sketch of controlling the MIT by ultrafast light pulses in magnetite. Two paths are generated from the initial state (equilibrium). Path 1 (2) corresponds to the process induced by the 800\,nm (400\,nm) and their effect on the electronic state, the trimerons arrangements, and the structure.}
\label{fig_SI:9} 
\end{center}
\end{figure}
\newpage
\bibliographystyle{unsrt} 
\bibliography{magnetiteSI}